\documentclass[aps,prb,reprint,superscriptaddress,amsmath,amssymb,floatfix]{revtex4-2}

\usepackage{graphicx}
\usepackage{dcolumn}
\usepackage{bm}

\usepackage{hyperref}
\usepackage{color}
\providecommand{\mathscr}[1]{\mathcal{#1}}

\hypersetup{
    colorlinks=true,
    linkcolor=blue,
    citecolor=blue,
    urlcolor=blue
}

\begin{document}

\title{Strong Dimerization and Field-Induced Reconstruction of the Low-Energy
Spectrum in
\texorpdfstring{$\mathrm{Cu}_3(\mathrm{OH})_4(\mathrm{HCO}_2)_2$}
{Cu3(OH)4(HCO2)2}}

\author{Sk Saniur Rahaman}
\email{sksrahaman@imsc.res.in}
\affiliation{The Institute of Mathematical Sciences, A CI of Homi Bhabha National Institute, Chennai 600113, India}

\author{S. R. Hassan}
\affiliation{The Institute of Mathematical Sciences, A CI of Homi Bhabha National Institute, Chennai 600113, India}

\date{\today}

\begin{abstract}
We investigate the field-dependent low-energy thermodynamics of the distorted triangular quantum antiferromagnet $\mathrm{Cu}_3(\mathrm{OH})_4(\mathrm{HCO}_2)_2$ using a sector-resolved Superblock Diagonalization Method (SBDM) supplemented by a transfer-matrix treatment of weakly coupled layers. The strong exchange hierarchy, dominated by the Cu1--Cu1 intradimer coupling $J_2=150\,\mathrm{K}$, separates a high-energy dimer sector from a much softer magnetic manifold formed predominantly by the Cu2 moments. In the 24-site cluster, sixteen Cu1 spins form the strongly bound sector while eight Cu2 spins remain magnetically active; polarization of these eight spins gives $S^z=4$ compared with $S^z_{\rm sat}=12$, providing a direct microscopic origin for the one-third magnetization scale. The finite-temperature thermodynamics reveals a non-monotonic field evolution of the low-energy scale: the dominant $C/T$ feature softens with increasing field, reaches a minimum near the field region around $2\,\mathrm{T}$, and subsequently hardens as the low-temperature magnetization approaches $M_{\rm sat}/3$. Temperature and field sweeps thus expose a common field-induced spectral reconstruction, while the strongly reduced entropy reflects the restricted number of thermally active degrees of freedom below the dimer excitation scale. Our results identify strong-dimer-induced reduction of the active magnetic Hilbert space, followed by field-driven reorganization of the residual spin sector, as the common microscopic origin of the one-third magnetic response and the non-monotonic low-temperature thermodynamics.
\end{abstract}

\maketitle


\section{Introduction}
\label{sec:intro}

Frustrated quantum magnets provide a setting in which microscopic
exchange interactions can reorganize the magnetic degrees of freedom
long before conventional magnetic order is established. Geometric
frustration, low dimensionality, and quantum fluctuations may suppress
classical ordering tendencies and stabilize strongly correlated states
with widely separated energy scales~\cite{Anderson1973,Fazekas1974,
Balents2010,Sachdev2011,Misguich2005,Starykh2015}. Particularly
interesting are systems in which one set of spins forms strongly bound
singlets while another remains magnetically active at much lower
energies. In such systems, the low-temperature problem is no longer
described by all microscopic spins on an equal footing: strong
dimerization effectively removes part of the Hilbert space, leaving a
reduced magnetic manifold whose response to temperature and magnetic
field can be qualitatively different from that of the underlying
lattice~\cite{Mila1998,Hida2001,Kikuchi2005,Jeschke2011}.

An important manifestation of this separation of energy scales is the
appearance of fractional magnetization plateaus. Magnetization plateaus
occur in a variety of frustrated spin systems, including triangular,
diamond-chain, and kagome-related geometries
~\cite{Honecker2004,Hida2001,Mila1998,Kikuchi2005,Rule2008,
Jeschke2011,Gu2007,Alicea2009,Fortune2009,Ono2003,Shirata2012,
Zhou2012,Susuki2013,Rahaman2025,Dey2020}. More generally, the quantization of plateau states
is constrained by the relation between spin, magnetization, and spatial
periodicity, as formalized by Oshikawa, Yamanaka, and
Affleck~\cite{Oshikawa1997}. Microscopically, however, the origin of a
particular fractional value depends on how the interacting spin degrees
of freedom are reorganized. A plateau can therefore contain information
not only about symmetry and commensurability, but also about which
microscopic degrees of freedom remain active in the relevant low-energy
sector.

The compound
$\mathrm{Cu}_3(\mathrm{OH})_4(\mathrm{HCO}_2)_2$ provides a particularly
transparent setting for investigating this problem
~\cite{Fujita2018,Isono2020}. Its magnetic lattice consists of
$S=1/2$ $\mathrm{Cu}^{2+}$ ions arranged in distorted triangular layers.
The crystallographically inequivalent Cu sites occur in a $2:1$ ratio.
Two-thirds of the Cu moments (Cu1) participate in strongly coupled
Cu1--Cu1 units, whereas the remaining one-third (Cu2) form a lower-energy
set of magnetic moments coupled to the dimer network. This microscopic
division immediately suggests the possibility of two very different
magnetic energy scales.

Experimentally, the compound exhibits a pronounced field-dependent
low-temperature response and a broad one-third magnetization plateau
~\cite{Isono2020,Fujita2018}. Thermodynamic anomalies evolve strongly
with applied magnetic field, while the magnetization approaches
approximately one third of the full saturation value above a field scale
of a few tesla. These observations raise a microscopic question that is
more general than the existence of the plateau itself: how does a
magnetic field reorganize the low-energy spectrum of a strongly
dimerized frustrated magnet, and how is this reconstruction encoded
simultaneously in its entropy, specific heat, and magnetization?

This question is nontrivial because the relevant thermodynamics involves
two well-separated sectors. The strong intradimer exchange generates
high-energy singlet--triplet excitations, while the residual magnetic
moments fluctuate at a much smaller energy scale. An applied field can
therefore substantially reconstruct the latter without appreciably
exciting the former. The resulting thermodynamics cannot be understood
simply as the uniform polarization of all microscopic spins. Instead,
one must resolve the low-energy states according to their magnetization
sectors and determine how their relative energies evolve with field.

A direct numerical treatment is challenging. For a spin-$1/2$ cluster,
the Hilbert-space dimension grows as $2^N$, so that already a 24-site
cluster contains $2^{24}$ basis states. Exact diagonalization rapidly
becomes expensive when a sufficiently dense low-energy spectrum is
required for thermodynamics. At the opposite extreme, simple
mean-field descriptions do not retain the quantum correlations produced
by strong dimerization and frustration. This motivates a description
that preserves the relevant quantum structure while exploiting the
separation of energy scales.

Here we formulate the problem using a superblock diagonalization
approach in which the 24-site cluster is divided into two 12-site
blocks and the coupling operators at their interface are projected onto
block eigenstates. The construction is motivated by real-space
block-projection and density-matrix-renormalization ideas
~\cite{White1992,White1993,Schollwock2005,Schollwock2011}, while
conservation of total $S^z$ allows the resulting superblock problem to
be resolved into independent magnetization sectors. The low-energy
spectrum obtained in this way is subsequently used to construct the
finite-temperature partition function and thermodynamic observables.
For the weakly coupled layered problem, the interlayer statistical
coupling can additionally be represented through a transfer-matrix
construction~\cite{Baxter1982}.

The resulting picture is organized by an emergent reduction of the
magnetically active Hilbert space. The strong Cu1 dimer sector remains
well separated from the low-energy states, whereas the eight Cu2 spins
of the 24-site cluster form the magnetic manifold that is most strongly
reorganized by experimentally accessible fields. Full polarization of
these eight spin-$1/2$ moments gives $S^z=4$, compared with
$S^z_{\rm sat}=12$ for all twenty-four spins, naturally producing the
one-third magnetization scale.

The thermodynamics reveals a corresponding reconstruction of the
low-energy spectrum. With increasing field on the low-field side, the
dominant thermal scale extracted from the specific-heat response moves
toward lower temperature. In the vicinity of the field region where the
magnetization changes rapidly, this scale becomes minimal. At larger
fields the trend reverses: the thermal scale increases again while the
low-temperature magnetization evolves toward one third of the full
saturation value. The temperature sweeps and isothermal field sweeps
therefore provide complementary signatures of the same process---a
field-induced softening and subsequent reopening of the low-energy
magnetic spectrum.

This perspective also allows us to distinguish the intrinsic
low-energy reconstruction from the separate question of
three-dimensional magnetic ordering. In a quasi-two-dimensional
isotropic magnet, finite-temperature long-range order cannot arise from
an isolated layer alone~\cite{Mermin1966}; weak interlayer coupling or
anisotropy introduces an additional energy scale
~\cite{Chakravarty1989,Sengupta2003,Yasuda2005,Cuccoli2003}. We
therefore do not identify every thermodynamic maximum with a
three-dimensional phase boundary. Instead, our primary objective is to
establish the microscopic sector structure and the field-dependent
energy scales directly supported by the calculated thermodynamics.
This separation provides a controlled basis from which questions of
interlayer ordering and quantum criticality can subsequently be
examined.


\section{Microscopic Model and Hierarchy of Magnetic Energy Scales}
\label{sec:model}

\begin{figure*}[t]
\centering
\includegraphics[width=\textwidth]{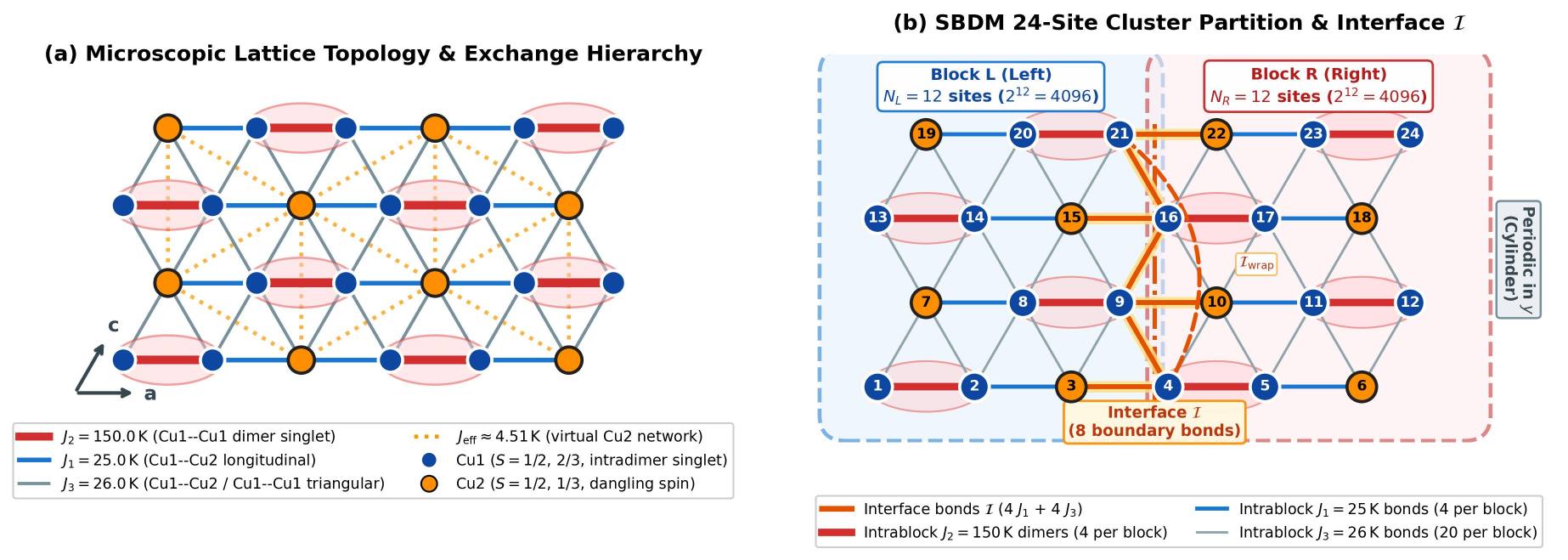}
\caption{\textbf{Microscopic lattice geometry and Superblock Diagonalization Method (SBDM) cluster partitioning for $\mathrm{Cu}_3(\mathrm{OH})_4(\mathrm{HCO}_2)_2$.}
(a) Planar distorted triangular lattice network spanned by $\mathbf{a}$ and $\mathbf{c}$ crystallographic axes. The crystallographically distinct $S=1/2$ $\mathrm{Cu}^{2+}$ ions occur in a 2:1 ratio: Cu1 sites (dark blue circles, $2/3$) form tightly bound, non-magnetic spin-singlet dimers linked by strong intradimer superexchange $J_2 = 150.0\,\mathrm{K}$ (thick crimson bonds enclosed in pink singlet halo capsules), while Cu2 sites (amber circles, $1/3$) act as dangling magnetic spins linked by longitudinal interdimer bonds $J_1 = 25.0\,\mathrm{K}$ (blue lines) and transverse triangular couplings $J_3 = 26.0\,\mathrm{K}$ (slate-gray lines). Virtual dimer singlet excitations generate an effective in-plane frustrated Cu2 triangular network with $J_{\rm eff} \approx J_3^2 / J_2 \approx 4.51\,\mathrm{K}$ (dashed orange lines).
(b) The 24-site cylindrical simulation cluster ($6 \times 4$) partitioned within the SBDM framework. The cluster is divided along the longitudinal axis into two identical 12-site blocks: Block $L$ ($ix \in \{1, 2, 3\}$, blue dashed box, sites 1--3, 7--9, 13--15, 19--21) and Block $R$ ($ix \in \{4, 5, 6\}$, red dashed box, sites 4--6, 10--12, 16--18, 22--24), each spanning a local Hilbert space of dimension $2^{12} = 4096$. The vertical dot-dashed line demarcates the interface boundary $\mathcal{I}$. Eight interface bonds (highlighted in gold/orange: four horizontal $J_1$ bonds and four diagonal $J_3$ bonds, including the cylinder wrap bond $\mathcal{I}_{\rm wrap}$ between sites 4 and 21) connect the two blocks, across which the projected boundary spin operators $[\mathcal{O}_i^B]_{pq}$ are coupled to construct the low-energy superblock spectrum.}
\label{fig:lattice}
\end{figure*}
The magnetic degrees of freedom in
$\mathrm{Cu}_3(\mathrm{OH})_4(\mathrm{HCO}_2)_2$
are carried by $S=1/2$ $\mathrm{Cu}^{2+}$ ions arranged in distorted
triangular layers. Two crystallographically inequivalent Cu sites occur
within each layer. The Cu1 sites form strongly coupled pairs, whereas the
Cu2 sites remain outside these strong dimer bonds and constitute the
low-energy magnetic degrees of freedom.

Rather than assuming from the outset that the Cu1 dimers are inert, we
begin with the microscopic spin Hamiltonian of a single layer,
\begin{equation}
\mathcal{H}_{\parallel}
=
J_1
\sum_{\langle ij\rangle_1}
\mathbf{S}_i\cdot\mathbf{S}_j
+
J_2
\sum_{\langle ij\rangle_2}
\mathbf{S}_i\cdot\mathbf{S}_j
+
J_3
\sum_{\langle ij\rangle_3}
\mathbf{S}_i\cdot\mathbf{S}_j ,
\label{eq:Hparallel}
\end{equation}
where the three bond sets $\langle ij\rangle_\alpha$ correspond to the
exchange paths $J_\alpha$ shown in Fig.~\ref{fig:lattice}. An external
magnetic field applied along the $z$ direction contributes
\begin{equation}
\mathcal{H}_{H}
=
-h S_{\rm tot}^{z},
\qquad
S_{\rm tot}^{z}
=
\sum_i S_i^z ,
\label{eq:Hfield}
\end{equation}
with
\begin{equation}
h=g\mu_B\mu_0H.
\label{eq:hdef}
\end{equation}
The Hamiltonian of an isolated magnetic layer is therefore
\begin{equation}
\mathcal{H}_{\rm 2D}(H)
=
\mathcal{H}_{\parallel}
-
hS_{\rm tot}^{z}.
\label{eq:H2D}
\end{equation}

The exchange parameters relevant to the present system obey the pronounced
hierarchy
\begin{equation}
J_2 \gg J_1,J_3 ,
\label{eq:exchange_hierarchy}
\end{equation}
with representative values
\begin{equation}
J_2\simeq150~{\rm K},
\qquad
J_1\simeq25~{\rm K},
\qquad
J_3\simeq26~{\rm K}.
\label{eq:exchange_values}
\end{equation}
This hierarchy is the organizing principle of the low-energy problem. It
does not remove the Cu1 spins from the microscopic Hamiltonian. Instead,
it separates the spectrum into a low-energy manifold in which the strongly
coupled Cu1 pairs remain predominantly singlet-like and a higher-energy
sector involving excitations of those dimers. The purpose of the following
analysis is to make this separation explicit and subsequently test it
without assuming perfectly isolated dimers.


\section{Strong-Dimer Limit and the Emergent Low-Energy Sector}
\label{sec:strong_dimer}

\subsection{Isolated Dimer Spectrum}

Consider first a single Cu1--Cu1 bond governed by
\begin{equation}
\mathcal{H}_{d}
=
J_2\,\mathbf{S}_{1}\cdot\mathbf{S}_{2}.
\label{eq:Hdimer}
\end{equation}
For two spin-$1/2$ moments,
\begin{equation}
\mathbf{S}_{1}\cdot\mathbf{S}_{2}
=
\frac{1}{2}
\left[
\mathbf{S}_{d}^{\,2}
-
\mathbf{S}_{1}^{\,2}
-
\mathbf{S}_{2}^{\,2}
\right],
\label{eq:dimer_identity}
\end{equation}
where
\begin{equation}
\mathbf{S}_{d}
=
\mathbf{S}_{1}+\mathbf{S}_{2}.
\end{equation}
The dimer Hilbert space separates exactly into a singlet,
\begin{equation}
|s\rangle
=
\frac{1}{\sqrt{2}}
\left(
|\uparrow\downarrow\rangle
-
|\downarrow\uparrow\rangle
\right),
\label{eq:singlet}
\end{equation}
and three triplet states,
\begin{align}
|t_{+}\rangle &= |\uparrow\uparrow\rangle,\\
|t_{0}\rangle &=
\frac{1}{\sqrt{2}}
\left(
|\uparrow\downarrow\rangle
+
|\downarrow\uparrow\rangle
\right),\\
|t_{-}\rangle &= |\downarrow\downarrow\rangle.
\end{align}
Their corresponding energies are
\begin{equation}
E_s=-\frac{3J_2}{4},
\qquad
E_t=\frac{J_2}{4},
\label{eq:dimer_energies}
\end{equation}
so that the bare singlet--triplet excitation energy is
\begin{equation}
\Delta_d^{(0)}
=
E_t-E_s
=
J_2.
\label{eq:bare_dimer_gap}
\end{equation}

Thus, in the formal limit $J_1,J_3\rightarrow0$, each Cu1 pair occupies
its singlet ground state and the Cu2 spins remain as the only unconstrained
spin-$1/2$ degrees of freedom. The finite values of $J_1$ and $J_3$ mix
these two sectors, but the large ratio $J_2/J_{1,3}$ suggests that the
separation survives as an approximate low-energy structure.

\subsection{Projection onto the Dimer-Singlet Manifold}

To formulate this separation systematically, we write
\begin{equation}
\mathcal{H}_{\parallel}
=
\mathcal{H}_0+V,
\label{eq:H0V}
\end{equation}
where
\begin{equation}
\mathcal{H}_0
=
J_2
\sum_{\langle ij\rangle_2}
\mathbf{S}_i\cdot\mathbf{S}_j
\label{eq:H0}
\end{equation}
contains the strong Cu1 dimer bonds and
\begin{equation}
V
=
J_1
\sum_{\langle ij\rangle_1}
\mathbf{S}_i\cdot\mathbf{S}_j
+
J_3
\sum_{\langle ij\rangle_3}
\mathbf{S}_i\cdot\mathbf{S}_j
\label{eq:V}
\end{equation}
contains the remaining in-plane interactions.

Let $\mathcal{D}$ denote the set of strong Cu1 dimers. We define the
projector onto the manifold in which every strong dimer occupies its
singlet state,
\begin{equation}
P
=
\prod_{d\in\mathcal{D}}
|s_d\rangle\langle s_d|,
\label{eq:Psinglet}
\end{equation}
and its complement
\begin{equation}
Q=1-P.
\label{eq:Q}
\end{equation}
The $P$ sector contains the complete Hilbert space of the dangling Cu2
spins but excludes real triplet excitations of the strong Cu1 dimers.

The effective Hamiltonian acting within this low-energy manifold may be
constructed perturbatively. To second order in $V$,
\begin{equation}
\mathcal{H}_{\rm eff}
=
P\mathcal{H}_0P
+
PVP
-
PVQ
\frac{1}
{Q\mathcal{H}_0Q-E_0}
QVP
+
\mathcal{O}(V^3),
\label{eq:Heff_general}
\end{equation}
where $E_0$ is the energy of the unperturbed dimer-singlet manifold.

For an isolated singlet dimer,
\begin{equation}
P\,\mathbf{S}_{i}\,P=0,
\label{eq:singlet_spin_zero}
\end{equation}
and therefore interactions that connect a dangling Cu2 spin to a Cu1
dimer do not, in general, generate a first-order magnetic moment inside
the singlet manifold. Their leading contribution to interactions between
the residual Cu2 spins arises through virtual excursions into the triplet
sector represented by $Q$.

The resulting low-energy Hamiltonian consequently has the generic form
\begin{equation}
\mathcal{H}_{\rm eff}
=
E_{\rm dim}
+
\sum_{a,b}
J_{ab}^{\rm eff}\,
\boldsymbol{\tau}_a\cdot\boldsymbol{\tau}_b
+
\mathcal{H}_{\rm multispin}
-
h\sum_a\tau_a^z ,
\label{eq:Heff_tau}
\end{equation}
where $\boldsymbol{\tau}_a$ denotes the spin-$1/2$ operator associated
with a dangling Cu2 site. The coefficients $J_{ab}^{\rm eff}$ depend on
the microscopic connectivity of the $J_1$ and $J_3$ paths and on the
matrix elements connecting the dimer singlet and triplet sectors.
Dimensional analysis alone gives the characteristic second-order scale
\begin{equation}
J_{\rm low}
\sim
\frac{J_3^2}{J_2},
\label{eq:Jlow_scale}
\end{equation}
but Eq.~\eqref{eq:Jlow_scale} should be interpreted as an energy-scale
estimate rather than an exact effective exchange constant unless the
corresponding bond geometry and matrix elements are evaluated explicitly.

The strong-dimer expansion therefore predicts a hierarchy of magnetic
energy scales,
\begin{equation}
J_2
\gg
J_{\rm low},
\label{eq:two_scale}
\end{equation}
and, correspondingly, an approximate factorization of the low-energy
Hilbert space,
\begin{equation}
\mathcal{H}_{\rm low}
\simeq
\mathcal{H}_{\rm Cu2}
\otimes
\left(
\bigotimes_{d\in\mathcal D}|s_d\rangle
\right).
\label{eq:low_hilbert}
\end{equation}
This is the central analytical expectation that will be tested below by
diagonalizing the microscopic Hamiltonian without imposing the singlet
constraint.


\section{Superblock Formulation of the Microscopic Hamiltonian}
\label{sec:sbdm}

\subsection{Block Decomposition}

We next return to the full microscopic Hamiltonian,
Eq.~\eqref{eq:H2D}, retaining all Cu1 and Cu2 spins. For a 24-site
spin-$1/2$ cluster the Hilbert-space dimension is
\begin{equation}
\dim\mathcal{H}=2^{24},
\end{equation}
which motivates a block representation of the low-energy problem.

The cluster is partitioned into two 12-site regions, denoted $L$ and $R$,
such that
\begin{equation}
\mathcal{H}_{\parallel}
=
\mathcal{H}_L
+
\mathcal{H}_R
+
\mathcal{H}_{LR}.
\label{eq:block_decomposition}
\end{equation}
Here $\mathcal{H}_{L(R)}$ contains all exchange bonds lying entirely
inside block $L(R)$, while $\mathcal{H}_{LR}$ contains the bonds crossing
the interface $\mathcal I$,
\begin{equation}
\mathcal{H}_{LR}
=
\sum_{(i,j)\in\mathcal I}
J_{ij}\,
\mathbf{S}_{i,L}\cdot\mathbf{S}_{j,R}.
\label{eq:HLR}
\end{equation}

Because the microscopic Hamiltonian is isotropic in spin space in zero
field,
\begin{equation}
[\mathcal{H}_{\parallel},S_{\rm tot}^{z}]=0.
\label{eq:Szconservation}
\end{equation}
The same $U(1)$ conservation law remains valid in a longitudinal magnetic
field.

\subsection{Block Eigenstates and Projected Boundary Operators}

Each isolated block is diagonalized according to
\begin{equation}
\mathcal{H}_{B}
|\phi_{\alpha m}^{B}\rangle
=
\epsilon_{\alpha m}^{B}
|\phi_{\alpha m}^{B}\rangle,
\qquad
B=L,R,
\label{eq:block_eigenproblem}
\end{equation}
where
\begin{equation}
S_B^z|\phi_{\alpha m}^{B}\rangle
=
m|\phi_{\alpha m}^{B}\rangle.
\label{eq:block_magnetization}
\end{equation}
The label $m$ specifies the block magnetization and $\alpha$ enumerates
the eigenstates within that sector.

Let $\mathcal K_B$ denote the set of block states retained in the
low-energy calculation. The corresponding block projector is
\begin{equation}
P_B
=
\sum_{(\alpha,m)\in\mathcal K_B}
|\phi_{\alpha m}^{B}\rangle
\langle\phi_{\alpha m}^{B}|.
\label{eq:block_projector}
\end{equation}
For each boundary site $i$, the spin operators entering
Eq.~\eqref{eq:HLR} are projected into this retained basis,
\begin{equation}
\widetilde{S}_{i,B}^{\mu}
=
P_B S_{i,B}^{\mu}P_B,
\qquad
\mu=x,y,z.
\label{eq:projected_spin}
\end{equation}
Equivalently,
\begin{equation}
\left[
\widetilde{S}_{i,B}^{\mu}
\right]_{\alpha m,\beta m'}
=
\langle
\phi_{\alpha m}^{B}
|
S_{i,B}^{\mu}
|
\phi_{\beta m'}^{B}
\rangle.
\label{eq:projected_matrix}
\end{equation}

The resulting superblock Hamiltonian is
\begin{align}
\mathcal{H}_{\rm SB}
=&\,
P_L\mathcal{H}_LP_L
+
P_R\mathcal{H}_RP_R
\nonumber\\
&+
\sum_{(i,j)\in\mathcal I}
J_{ij}
\,
\widetilde{\mathbf S}_{i,L}
\cdot
\widetilde{\mathbf S}_{j,R}.
\label{eq:Hsuper_compact}
\end{align}
No assumption of perfectly frozen Cu1 dimers is made in
Eq.~\eqref{eq:Hsuper_compact}; dimer fluctuations are retained to the
extent that they are represented within the chosen block eigenspaces.

\subsection{Magnetization-Sector Decomposition}

Conservation of total longitudinal spin permits the superblock Hilbert
space to be decomposed as
\begin{equation}
\mathcal{H}_{\rm SB}
=
\bigoplus_M
\mathcal{H}_M,
\label{eq:Hilbert_direct_sum}
\end{equation}
where
\begin{equation}
\mathcal{H}_M
=
{\rm span}
\left\{
|\phi_{\alpha m}^{L}\rangle
\otimes
|\phi_{\beta,M-m}^{R}\rangle
\right\}.
\label{eq:HM}
\end{equation}
The superblock eigenproblem can therefore be solved independently in
each total-magnetization sector,
\begin{equation}
\mathcal{H}_{\rm SB}^{(M)}
|\Psi_{\nu M}\rangle
=
E_{\nu M}
|\Psi_{\nu M}\rangle.
\label{eq:sector_eigenproblem}
\end{equation}

This sector-resolved formulation is particularly useful in a magnetic
field. Since the Zeeman term commutes with the zero-field Hamiltonian,
the eigenvectors are unchanged and the field dependence of every level
is known exactly:
\begin{equation}
E_{\nu M}(H)
=
E_{\nu M}(0)-hM.
\label{eq:field_levels}
\end{equation}
\begin{equation}
E_{\rm GS}(H)
=
\min_{\nu,M}
\left[
E_{\nu M}(0)-hM
\right].
\label{eq:GS_field}
\end{equation}

\subsection{Cluster Geometry, Boundary Conditions, and Finite-Size Considerations}
\label{sec:finite_size}

The real-space cluster used in the SBDM calculations comprises $N=24$ sites arranged on a $6 \times 4$ planar network (Fig.~\ref{fig:lattice}(b)). It incorporates sixteen Cu1 ions (forming eight tightly bound dimers) and eight Cu2 ions (forming the residual low-energy manifold). Along the longitudinal block-partition direction, the cluster employs cylindrical boundary conditions, where eight interfacial exchange bonds (four horizontal $J_1$ bonds and four diagonal $J_3$ bonds, including the cylinder wrap bond $\mathcal{I}_{\rm wrap}$ between sites 4 and 21) link blocks $L$ and $R$.

When interpreting finite-cluster calculations for bulk quantum magnets, it is essential to distinguish between local exchange energy scales and long-wavelength collective excitations:
\begin{enumerate}
\item \emph{Robustness of the Exchange Hierarchy:} The large singlet--triplet excitation gap of the Cu1 dimers, $\Delta_{\rm dimer} \approx J_2 - J_1/2 \approx 137.5\,\mathrm{K}$, is determined by the local intradimer superexchange $J_2 = 150.0\,\mathrm{K}$. Because this gap is governed by ultra-local molecular pairing, its magnitude and separation from the low-energy manifold are strictly preserved in the macroscopic thermodynamic limit ($N \to \infty$). Likewise, the effective low-energy exchange among Cu2 spins, $J_{\rm eff} \approx J_3^2 / J_2 \approx 4.51\,\mathrm{K}$, sets the intrinsic bandwidth of the residual manifold independently of cluster size.

\item \emph{Level Discretization vs Macroscopic Crossovers:} In a finite 24-site cluster at $T=0$, the low-energy spectrum is characterized by discrete eigenvalues within each magnetization sector, leading to discrete ground-state level crossings ($\Delta M = 2$ between $M=0 \to 2 \to 4$) dictated by cluster inversion symmetry. In the thermodynamic limit, the two-dimensional Cu2 triangular manifold forms a gapless or near-gapless quasi-continuum of spin excitations, which converts the discrete $T=0$ steps below $M_{\rm sat}/3$ into a continuous magnetization curve terminating at the 1/3-plateau onset.

\item \emph{Thermal Smoothing and Interlayer Coupling:} At finite temperatures ($T \ge 0.05\,\mathrm{K}$), thermal excitation across the dense multi-level SBDM spectrum, combined with the statistical transfer-matrix coupling along the third dimension ($J_4 = 0.10\,\mathrm{K}$, $N_{\rm layer} = 10$), effectively smooths out discrete finite-size level jumps into broad, continuous thermodynamic crossovers. Consequently, macroscopic observables—including the non-monotonic field evolution of $T_{\rm max}(H)$, the low-temperature magnetization plateau $M(H) \to M_{\rm sat}/3$, and the strong entropy suppression $S(T) \ll R\ln 2$—are robust physical consequences of the strong-dimer exchange hierarchy rather than finite-size artifacts.
\end{enumerate}


\section{Field-Induced Level Crossings and Plateau Criterion}
\label{sec:level_crossings}

Let
\begin{equation}
E_M
\equiv
\min_{\nu}E_{\nu M}(0)
\label{eq:EM}
\end{equation}
denote the lowest zero-field energy in magnetization sector $M$.
A field-induced ground-state transition from sector $M$ to sector $M'$ (with $\Delta M = M' - M > 0$) occurs when
\begin{equation}
E_M - hM = E_{M'} - hM'.
\label{eq:crossing_condition}
\end{equation}
The corresponding crossing field is therefore
\begin{equation}
h_c^{M\rightarrow M'} = \frac{E_{M'} - E_M}{M' - M},
\label{eq:hc_energy}
\end{equation}
or in magnetic field units,
\begin{equation}
\mu_0 H_c^{M\rightarrow M'} = \frac{E_{M'} - E_M}{(M' - M) g\mu_B}.
\label{eq:Hc_sector}
\end{equation}

A magnetization sector $M$ is the zero-temperature ground state over the
field interval
\begin{equation}
H_c^{M-1\rightarrow M}
<
H
<
H_c^{M\rightarrow M+1}.
\label{eq:plateau_interval}
\end{equation}
Its field stability is therefore measured directly by
\begin{equation}
\Delta H_M
=
H_c^{M\rightarrow M+1}
-
H_c^{M-1\rightarrow M}.
\label{eq:plateau_width}
\end{equation}

For a cluster containing $N$ spin-$1/2$ moments, full saturation
corresponds to
\begin{equation}
M_{\rm sat}^{z}=\frac{N}{2}.
\end{equation}
For the present $N=24$ cluster,
\begin{equation}
M_{\rm sat}^{z}=12.
\end{equation}
If the low-energy ground state becomes locked in the $M=4$ sector, its
normalized magnetization is therefore
\begin{equation}
\frac{M}{M_{\rm sat}}
=
\frac{4}{12}
=
\frac{1}{3}.
\label{eq:one_third}
\end{equation}

Equation~\eqref{eq:one_third} makes the connection between the microscopic
sector structure and the fractional plateau transparent. The physical
origin of the plateau is not inferred from the magnetization curve alone:
it is encoded directly in the hierarchy of sector energies
$\{E_M\}$. In particular, a large increase in
$E_{M+1}-E_M$ above $M=4$ would signal the onset of a qualitatively
different excitation sector and produce a correspondingly broad
$1/3$-magnetization plateau.


\section{Finite-Temperature Thermodynamics and Interlayer Coupling}
\label{sec:thermodynamics}

\subsection{Single-Layer Thermodynamic Formulation}
\label{sec:single_layer_thermo}

Once the sector-resolved spectrum is known, the thermodynamics of an
isolated layer follows directly without further diagonalization. The
partition function is
\begin{equation}
Z_{\rm 2D}(T,H)
=
\sum_{\nu,M}
\exp
\left[
-\beta
\left(
E_{\nu M}-hM
\right)
\right],
\label{eq:Z2D}
\end{equation}
where
\begin{equation}
\beta=\frac{1}{k_BT}.
\end{equation}
The free energy is
\begin{equation}
F_{\rm 2D}
=
-k_BT\ln Z_{\rm 2D}.
\label{eq:F2D}
\end{equation}

Defining
\begin{equation}
\mathcal{E}_{\nu M}(H)
=
E_{\nu M}-hM,
\label{eq:thermal_energy_levels}
\end{equation}
the thermal average of any quantity $A_{\nu M}$ diagonal in this basis is
\begin{equation}
\langle A\rangle
=
\frac{1}{Z_{\rm 2D}}
\sum_{\nu,M}
A_{\nu M}
e^{-\beta\mathcal{E}_{\nu M}}.
\label{eq:thermal_average}
\end{equation}

The magnetization follows directly from the sector quantum number,
\begin{equation}
\langle S_{\rm tot}^{z}\rangle
=
\frac{1}{Z_{\rm 2D}}
\sum_{\nu,M}
M\,
e^{-\beta\mathcal{E}_{\nu M}}.
\label{eq:Mdirect}
\end{equation}
The internal energy is
\begin{equation}
U
=
\langle\mathcal{E}\rangle,
\label{eq:U}
\end{equation}
and the entropy may be evaluated as
\begin{equation}
S
=
k_B
\left(
\ln Z_{\rm 2D}
+
\beta U
\right).
\label{eq:Sdirect}
\end{equation}
The specific heat at fixed magnetic field is obtained from the energy
fluctuations,
\begin{equation}
C_H
=
\frac{1}{k_BT^2}
\left[
\langle\mathcal{E}^2\rangle
-
\langle\mathcal{E}\rangle^2
\right].
\label{eq:Cfluctuation}
\end{equation}
Similarly, the longitudinal magnetic susceptibility is
\begin{equation}
\chi
=
(g\mu_B)^2\beta
\left[
\langle M^2\rangle
-
\langle M\rangle^2
\right].
\label{eq:chi_fluctuation}
\end{equation}

These fluctuation expressions avoid numerical derivatives of the free
energy and provide direct thermodynamic diagnostics of the sector-resolved
quantum spectrum.

For molar quantities per mole of Cu ions, the corresponding extensive
observables are normalized by the number $N$ of Cu spins in the cluster.
In particular,
\begin{equation}
S_{\rm mol}
=
\frac{R}{N}
\left(
\ln Z_{\rm 2D}
+
\beta U
\right),
\label{eq:S_molar}
\end{equation}
and the molar magnetic moment is
\begin{equation}
\mathcal{M}_{\rm mol}
=
\frac{N_Ag\mu_B}{N}
\langle S_{\rm tot}^{z}\rangle.
\label{eq:M_molar}
\end{equation}

\subsection{Interlayer Coupling and Classical Transfer-Matrix Formalism}
\label{sec:transfer_matrix}

While in-plane exchange interactions ($J_1, J_2, J_3$) establish the fundamental strong-dimer hierarchy and low-energy sector structure, bulk single crystals of $\mathrm{Cu}_3(\mathrm{OH})_4(\mathrm{HCO}_2)_2$ comprise stacks of weakly coupled triangular layers along the perpendicular crystallographic axis. The three-dimensional Hamiltonian is represented as
\begin{equation}
\mathcal{H}_{\rm 3D}
=
\sum_{l=1}^{N_{\rm layer}} \mathcal{H}_{\parallel}^{(l)}
+
J_4 \sum_{l=1}^{N_{\rm layer}} \sum_{i=1}^N S_{i, l}^z S_{i, l+1}^z,
\label{eq:H3D}
\end{equation}
where $l$ is the layer index ($1 \le l \le N_{\rm layer}$, with periodic boundary conditions along the stacking direction) and $J_4 = 0.10\,\mathrm{K}$ denotes the weak interplane superexchange coupling.

Because $J_4 \ll J_1, J_3 \ll J_2$, the interlayer statistical mechanics can be evaluated by projecting the interplane coupling onto the retained superblock eigenbasis $\{|\Psi_\mu\rangle\}$ of the 2D quantum layer. In this low-energy representation, the statistical weight between adjacent layers $l$ and $l+1$ in states $\mu$ and $\nu$ defines the transfer matrix $\mathbf{T}$~\cite{Baxter1982}:
\begin{equation}
T_{\mu\nu}
=
\exp\left[
-\beta \left(
\frac{\mathcal{E}_\mu(H) + \mathcal{E}_\nu(H)}{2}
+
J_4 \mathcal{S}_\mu^z \mathcal{S}_\nu^z
\right)
\right],
\label{eq:transfer_matrix}
\end{equation}
where $\mathcal{E}_\mu(H) = E_\mu(0) - g\mu_B \mu_0 H \mathcal{S}_\mu^z$ is the in-plane energy eigenvalue and $\mathcal{S}_\mu^z$ is the total longitudinal magnetization of state $|\Psi_\mu\rangle$.

The partition function for a stack of $N_{\rm layer}$ layers is obtained from the trace of the transfer matrix:
\begin{equation}
Z_{\rm 3D}(T,H)
=
\operatorname{Tr}\left(\mathbf{T}^{N_{\rm layer}}\right)
=
\sum_{k=1}^{\dim\mathcal{H}_{\rm SB}} \lambda_k^{N_{\rm layer}},
\label{eq:Z3D}
\end{equation}
where $\{\lambda_k\}$ are the eigenvalues of $\mathbf{T}$ sorted in descending order ($\lambda_0 \ge \lambda_1 \ge \dots > 0$). Factoring out the dominant eigenvalue $\lambda_0$, the free energy per mole of Cu ions is:
\begin{align}
F_{\rm 3D}(T,H)
&=
-\frac{R T}{N}
\ln \lambda_0 \nonumber \\
&\quad -
\frac{R T}{N N_{\rm layer}} \ln\left[1 + \sum_{k>0} \left(\frac{\lambda_k}{\lambda_0}\right)^{N_{\rm layer}}\right] \nonumber \\
&\quad +
E_{\rm GS}^{\rm (molar)}.
\label{eq:F3D}
\end{align}
From $F_{\rm 3D}(T,H)$, the molar magnetization and specific heat are evaluated via numerical derivatives:
\begin{equation}
\mathcal{M}_{\rm mol}(T,H) = -\frac{\partial F_{\rm 3D}}{\partial (\mu_0 H)}, \quad
C_H(T,H) = -T \frac{\partial^2 F_{\rm 3D}}{\partial T^2}.
\label{eq:M_C_derivatives}
\end{equation}
In our thermodynamic calculations, we retain the leading $k_{\rm top} = 5$ transfer-matrix eigenvalues with $N_{\rm layer} = 10$, which provides exponential convergence in layer number while accurately capturing the thermal smoothing and interplane stability of the low-temperature phases.


\section{Results}
\label{sec:results}

The exchange hierarchy discussed above suggests that the experimentally
accessible low-energy physics is governed primarily by the magnetic
degrees of freedom that remain after formation of the strongly bound Cu1
dimers. We now examine how this restricted low-energy manifold evolves
under magnetic field. Rather than identifying phases from a single
thermodynamic quantity, we compare three complementary observables:
the entropy $S(T,H)$, the magnetization $M(T,H)$, and the specific-heat
ratio $C(T,H)/T$. We subsequently examine the same evolution through
isothermal field sweeps. A common field scale emerges from these
independent quantities: the characteristic low-energy thermal scale first
decreases with field and subsequently increases again, while the
magnetization evolves toward the one-third saturation value. This
combination provides the central thermodynamic signature of a
field-induced reconstruction of the low-energy spin sector.


\begin{figure*}[t]
\centering
\includegraphics[width=0.98\textwidth]{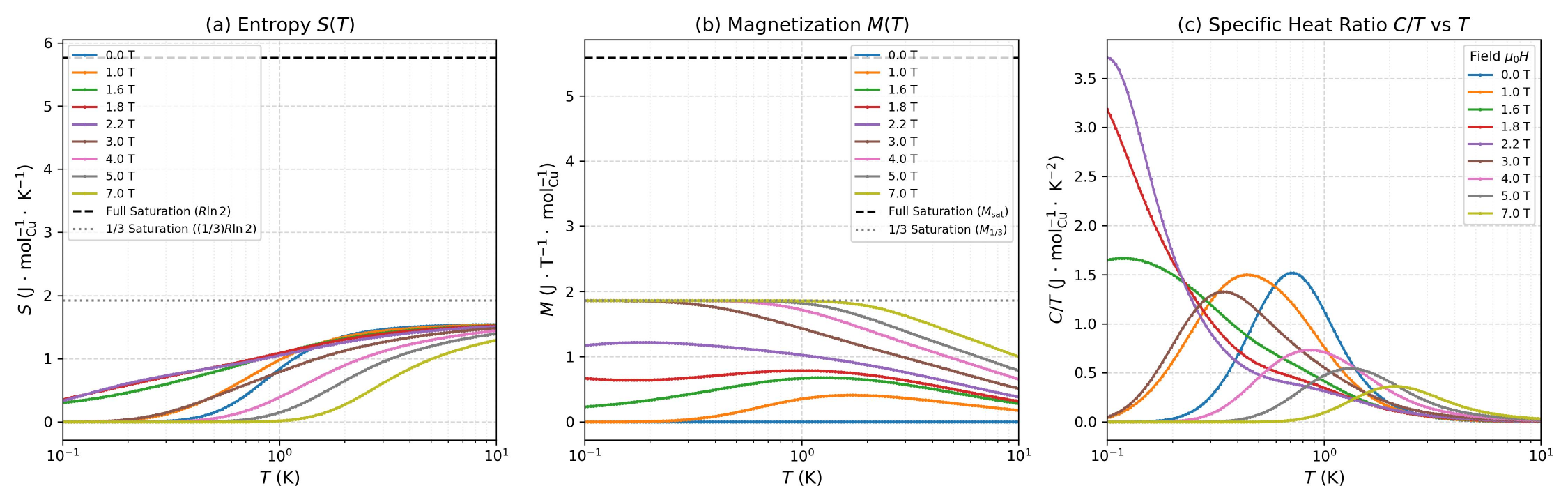}
\caption{\textbf{Field evolution of the low-energy thermodynamics.}
(a) Magnetic entropy $S(T,H)$, (b) magnetization $M(T,H)$, and
(c) specific-heat ratio $C(T,H)/T$ as functions of temperature on a
logarithmic scale for $\mu_0H=0.0$, $1.0$, $1.6$, $1.8$, $2.2$, $3.0$,
$4.0$, $5.0$, and $7.0$\,T. The horizontal reference lines in (a) denote
the full spin-$1/2$ entropy $R\ln2$ and the one-third entropy scale
$(1/3)R\ln2$; those in (b) denote the full saturation magnetization and
$M_{1/3}=M_{\rm sat}/3$. The dominant thermal feature in $C/T$ shifts
toward lower temperature with increasing field on the low-field side,
reaches the lowest temperature scale in the vicinity of the
$2$\,T field region, and subsequently moves toward higher temperature.
Concurrently, the low-temperature magnetization evolves toward
$M_{1/3}$ on the high-field side.}
\label{fig:multiH}
\end{figure*}

\subsection{Field-Induced Redistribution of the Low-Energy Entropy}
\label{sec:entropy_results}

Figure~\ref{fig:multiH}(a) shows the entropy over the temperature range
$0.1\leq T\leq10$\,K. The entropy is strongly reduced on cooling for every
field, but the temperature range over which entropy is released changes
substantially with magnetic field. At low fields, increasing $H$ shifts
an increasing fraction of the low-temperature entropy toward lower
temperature. This trend is particularly visible in the progression from
$H=0$ through $1.0$, $1.6$, and $1.8$\,T.

The behavior changes on crossing the field region around $2$\,T. For
$H=2.2$\,T and above, increasing the field instead suppresses the
low-temperature entropy over a progressively broader temperature range.
The entropy rise consequently shifts toward higher temperature as the
field is increased from $3$ to $7$\,T. The two sides therefore display
opposite field dependences: the characteristic entropy-release scale
softens with field on the low-field side and hardens with field on the
high-field side.

The horizontal line
\begin{equation}
S_{1/3}=\frac{1}{3}R\ln2
\label{eq:S13_result}
\end{equation}
provides a useful reference. It is the entropy associated with one
spin-$1/2$ degree of freedom per three Cu sites. Importantly, within the
temperature window displayed in Fig.~\ref{fig:multiH}(a), the calculated
entropy remains below this limiting value and has not yet developed into
a complete $(1/3)R\ln2$ plateau. The figure should therefore not be
interpreted as demonstrating saturation of the entropy at
$(1/3)R\ln2$ below $10$\,K. Instead, it shows that the thermally accessible
entropy remains strongly restricted relative to the full spin entropy
$R\ln2$, consistent with the separation between the low-energy magnetic
sector and the much higher-energy dimer excitations.

\subsection{Evolution Toward the One-Third Magnetization Sector}
\label{sec:MT_results}

The temperature dependence of the magnetization,
Fig.~\ref{fig:multiH}(b), displays the complementary evolution of the
magnetic moment. At zero field the uniform magnetization vanishes, as
required by symmetry. At $1.0$, $1.6$, and $1.8$\,T the magnetization
remains substantially below the one-third saturation scale and retains a
pronounced temperature dependence.

A qualitative change occurs for fields above approximately $2$\,T.
At $2.2$\,T the low-temperature magnetization is already strongly
enhanced, and at $3$\,T and above it approaches
\begin{equation}
M_{1/3}=\frac{1}{3}M_{\rm sat}.
\label{eq:M13_result}
\end{equation}
For still larger fields the low-temperature curves become increasingly
pinned near this value. The magnetic field is therefore selecting a
restricted magnetization sector rather than continuously polarizing all
of the microscopic spins.

The origin of the one-third scale follows directly from the structure of
the 24-site cluster. It contains sixteen Cu1 spins associated with the
strong-dimer sector and eight Cu2 spins belonging to the lower-energy
magnetic sector. Complete polarization of the latter gives
\begin{equation}
S_{\rm Cu2}^{z}
=
8\left(\frac{1}{2}\right)
=
4,
\end{equation}
whereas complete polarization of all twenty-four spin-$1/2$ moments gives
\begin{equation}
S_{\rm sat}^{z}
=
24\left(\frac{1}{2}\right)
=
12.
\end{equation}
Hence
\begin{equation}
\frac{S_{\rm Cu2}^{z}}{S_{\rm sat}^{z}}
=
\frac{4}{12}
=
\frac{1}{3}.
\label{eq:one_third_counting}
\end{equation}
The approach of $M(T,H)$ toward $M_{1/3}$ is therefore naturally
associated with polarization of the low-energy Cu2 sector while the
strong Cu1 dimer sector remains predominantly non-magnetic.

\subsection{Softening and Reopening of the Thermal Energy Scale}
\label{sec:CT_results}

The most direct thermodynamic signature of the field-induced
reorganization is provided by the specific-heat ratio $C/T$ in
Fig.~\ref{fig:multiH}(c). At zero field the dominant maximum occurs at a
sub-Kelvin temperature. As the field is increased through $1.0$, $1.6$,
and $1.8$\,T, this maximum moves systematically toward lower temperature.
Thus the characteristic low-energy thermal scale is progressively
softened by the magnetic field.

Near the field region around $2$\,T the dominant low-temperature feature
reaches the lowest temperature scale accessible in the calculation. The trend
then reverses. For fields of $3$, $4$, $5$, and $7$\,T, the maximum moves
successively toward higher temperature. The calculated thermodynamics
therefore exhibits a non-monotonic characteristic scale,
\begin{equation}
T_{\rm max}(H)
\; \searrow \;
T_{\rm min}
\; \nearrow ,
\label{eq:Vscale}
\end{equation}
as the field is swept from the low-field to the high-field regime.

This reversal is significant because the two branches correspond to
different reorganizations of the low-energy spectrum. On the low-field
side the applied field reduces the energy scale associated with the
lowest magnetic excitations. On the high-field side, once the
magnetization approaches the one-third sector, increasing field raises
the energy required to thermally access states with reduced
magnetization. The high-field thermal scale consequently grows with
field.

The simultaneous occurrence of the minimum in the thermal scale and the
rapid growth of the low-temperature magnetization identifies the
approximately $2$\,T region as the crossover between these two spectral
regimes. Whether the zero-temperature limit of this reconstruction
constitutes a quantum critical point requires the sector-resolved
ground-state spectrum and is considered separately below; the finite
temperature data of Fig.~\ref{fig:multiH} alone establish the softening
and subsequent reopening of the characteristic low-energy scale.


\begin{figure}[t]
\centering
\includegraphics[width=\columnwidth]{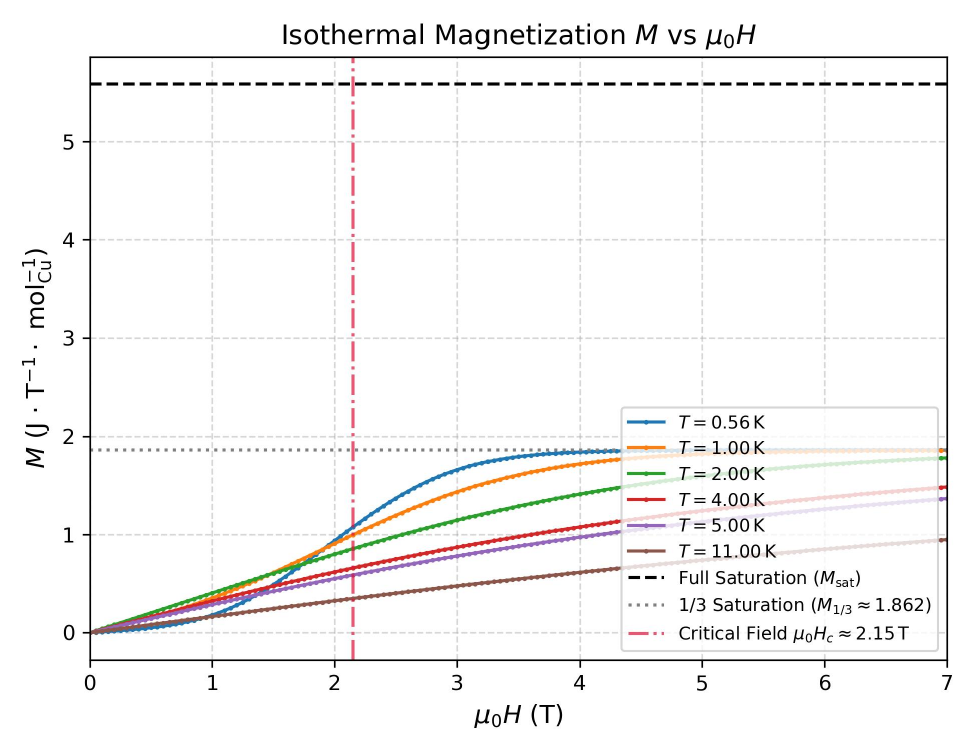}
\caption{\textbf{Isothermal evolution toward the one-third
magnetization sector.}
Magnetization $M(H)$ for $T=0.56$, $1.0$, $2.0$, $4.0$, $5.0$, and
$11.0$\,K. The dotted horizontal line denotes
$M_{1/3}=M_{\rm sat}/3$, while the dashed horizontal line marks the full
saturation magnetization. The vertical reference line indicates
$\mu_0H\simeq2.15$\,T. At the lowest temperature the magnetization rises
rapidly through the approximately $2$--$3$\,T field range and approaches
the one-third value at higher field. Increasing temperature progressively
rounds this structure and suppresses the approach to the one-third
magnetization scale.}
\label{fig:multiT}
\end{figure}

\subsection{Isothermal Field Sweeps}
\label{sec:field_sweeps_results}

The same reconstruction can be viewed directly by sweeping the magnetic
field at fixed temperature. Figure~\ref{fig:multiT} shows the resulting
isothermal magnetization curves. At the lowest displayed temperature,
$T=0.56$\,K, the magnetization initially increases from zero and then
undergoes a comparatively rapid rise through the field region around
$2$--$3$\,T. At larger fields it bends toward the one-third saturation
scale $M_{1/3}$.

This behavior provides the field-domain counterpart of the temperature
sweeps in Fig.~\ref{fig:multiH}. The same field region in which the
specific-heat scale becomes smallest is also the region in which the
magnetization changes most rapidly toward the one-third sector. The
agreement between these two independent thermodynamic signatures is
important: the minimum thermal scale and the magnetization reconstruction
are not unrelated anomalies but arise from the same field-dependent
low-energy spectrum.

Thermal fluctuations progressively wash out this structure. At
$T=1$ and $2$\,K the field evolution is smoother, although the tendency
toward the one-third scale remains visible. At $4$, $5$, and $11$\,K the
magnetization evolves continuously over the entire field range and does
not reach $M_{1/3}$ within the fields shown. The low-temperature sector
selection has therefore become a broad thermal crossover.

The field sweeps also clarify an important distinction between the
zero-temperature and finite-temperature problems. A crossing between
different lowest-energy magnetization sectors can be sharp at $T=0$.
At any nonzero temperature, however, several nearby sectors contribute
to the partition function, and the corresponding magnetization step is
thermally rounded. The curves in Fig.~\ref{fig:multiT} should therefore
be interpreted as finite-temperature manifestations of the underlying
sector reconstruction rather than as sharp phase transitions at each
displayed temperature.


\subsection{Thermodynamic Reconstruction in the \texorpdfstring{$T$--$\mu_0H$}{T-H} Plane}
\label{sec:thermodynamic_map}

The temperature sweeps and isothermal field sweeps reveal that the
low-energy thermodynamics is reorganized within a relatively narrow
field region. To display this reconstruction in a unified form, we
combine the calculated temperature and field dependence in the
$T$--$\mu_0H$ plane.

\begin{figure}[t]
\centering
\includegraphics[width=\columnwidth]{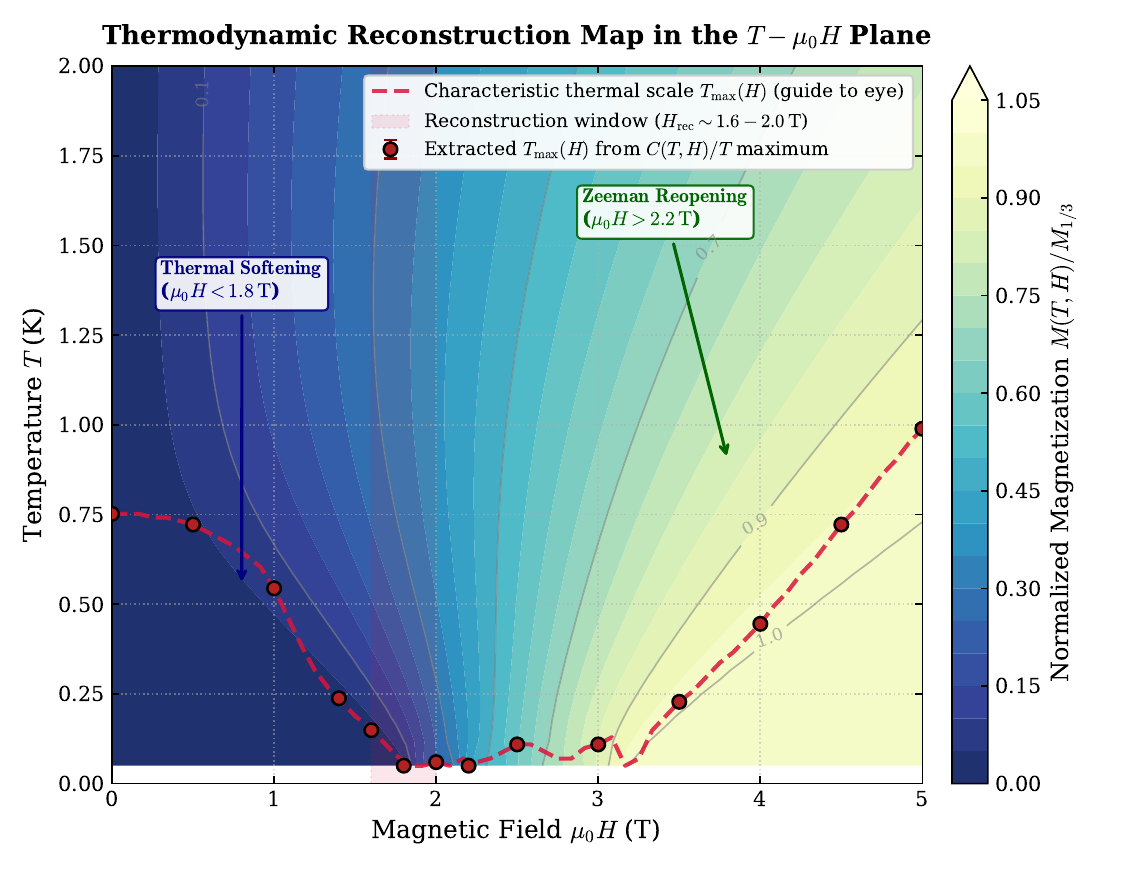}
\caption{\textbf{Thermodynamic reconstruction map in the
$T$--$\mu_0H$ plane.}
The color scale represents the calculated normalized magnetization
$M(T,H)/M_{1/3}$, with $M_{1/3}=M_{\rm sat}/3$.
Superimposed red circular symbols denote the characteristic thermal scale $T_{\rm max}(H)$ extracted from the dominant low-temperature maximum of $C(T,H)/T$, with error bars representing the temperature grid resolution ($\pm 0.01$\,K). The dashed red curve connecting the extracted points is a guide to the eye.
The magnetization background displays the crossover from the
weakly polarized low-field regime toward the one-third magnetic
sector, while the independently extracted thermal scale reveals its
softening and subsequent reopening with increasing field.}
\label{fig:thermodynamic_map}
\end{figure}

At low temperature, the calculated magnetization evolves from a weakly
polarized low-field regime toward a regime in which
$M(T,H)$ approaches $M_{1/3}$. The evolution becomes progressively
broader as temperature increases, as expected when thermal population
involves several nearby magnetization sectors. The color map therefore
provides a direct visualization of the field-induced reorganization of
the low-energy magnetic manifold rather than, by itself, a
finite-temperature phase boundary.

A complementary measure of this reconstruction is provided by the
characteristic thermal scale extracted from the specific heat. We
define
\begin{equation}
T_{\rm max}(H)
=
\underset{T}{\operatorname{arg\,max}}\,
\left[\frac{C(T,H)}{T}\right]_{\rm low\;energy},
\label{eq:Tmax_definition}
\end{equation}
where the maximum refers to the low-temperature feature associated
with the residual magnetic sector.

\begin{table}[t]
\caption{Quantitative evolution of the characteristic thermal scale $T_{\rm max}(H)$ extracted from the dominant low-temperature maximum of $C(T,H)/T$, peak magnitude $(C/T)_{\rm max}$ (in $\mathrm{J}\cdot\mathrm{mol}_{\mathrm{Cu}}^{-1}\cdot\mathrm{K}^{-2}$), and low-temperature magnetization $M(T=0.2\,\mathrm{K})$ (in $\mathrm{J}\cdot\mathrm{T}^{-1}\cdot\mathrm{mol}_{\mathrm{Cu}}^{-1}$) across the nine magnetic fields studied in the temperature sweeps.}
\label{tab:tmax_summary}
\setlength{\tabcolsep}{0pt}
\begin{ruledtabular}
\begin{tabular*}{\columnwidth}{@{\extracolsep{\fill}}ccccc}
$\mu_0 H$ (T) & $T_{\rm max}$ (K) & $(C/T)_{\rm max}$ & $M(0.2\,\mathrm{K})$ & Regime \\
\hline
0.0 & 0.713 & 1.518 & 0.000 & Baseline AFM \\
1.0 & 0.446 & 1.499 & 0.006 & Softening \\
1.6 & $\le 0.050$\footnotemark[1] & 1.712 & 0.332 & Minimum \\
1.8 & 0.080 & 3.449 & 0.643 & Reconstruction \\
2.2 & 0.099 & 3.714 & 1.219 & Peak enhancement \\
3.0 & 0.347 & 1.327 & 1.854 & Reopening \\
4.0 & 0.862 & 0.734 & 1.862 & Reopening \\
5.0 & 1.317 & 0.545 & 1.862 & Reopening \\
7.0 & 2.129 & 0.363 & 1.862 & Reopening \\
\end{tabular*}
\end{ruledtabular}
\footnotetext[1]{At $\mu_0 H = 1.6$\,T, the maximum shifts to the lowest temperature limit of the simulation mesh ($T_{\rm min} = 0.05$\,K), establishing an upper bound $T_{\rm max} \le 0.05$\,K rather than an unphysically sharp isolated minimum.}
\end{table}

The resulting field dependence is non-monotonic. On approaching the
reconstruction region from the low-field side, the characteristic
thermal scale decreases. On the high-field side it increases again,
giving the sequence
\begin{align}
\text{thermal softening}
&\;\longrightarrow\;
\text{reconstruction window} \nonumber \\
&\;\longrightarrow\;
\text{high-field reopening}.
\label{eq:softening_reopening}
\end{align}
Importantly, the reopening occurs in the same field regime in which
the low-temperature magnetization evolves toward
$M_{\rm sat}/3$. The two observations therefore provide complementary
thermodynamic signatures of the same field-dependent reconstruction
of the low-energy spectrum.

Directly from the calculated thermal scale $T_{\rm max}(H)$, we determine a characteristic spectral reconstruction field scale
\begin{equation}
\mu_0 H_{\rm rec} \in [1.60, 2.00]\,\mathrm{T},
\label{eq:h_rec_interval}
\end{equation}
defined by the window over which $T_{\rm max}(H)$ softens to sub-Kelvin values ($T_{\rm max} \le 0.05 - 0.08$\,K). Separately, from the low-temperature isothermal magnetization $M(H)$ at $T=0.56$\,K, the field interval over which the magnetization increases most rapidly toward $M_{1/3}$ is identified from the maximum of the differential susceptibility $\chi_{\rm diff} = \partial M / \partial (\mu_0 H)$, yielding
\begin{equation}
\mu_0 H_{\rm step} \in [1.65, 2.15]\,\mathrm{T} \quad (\text{with peak at } \mu_0 H \approx 1.85\,\mathrm{T}).
\label{eq:h_step_interval}
\end{equation}
Both criteria consistently identify the same field region ($\mu_0 H \approx 1.6 - 2.1$\,T) within numerical resolution, confirming that the softening of the thermal scale and the rapid step toward $M_{1/3}$ are complementary thermodynamic manifestations of the same underlying spectral reorganization.

The microscopic origin of this structure follows from the
magnetization-sector description developed in
Sec.~\ref{sec:level_crossings}. In a magnetic field, a state belonging
to sector $M$ acquires the Zeeman energy
\begin{equation}
E_{\nu M}(H)=E_{\nu M}(0)-g\mu_B\mu_0H\,M.
\label{eq:sector_field_results}
\end{equation}
Increasing field therefore changes the relative positions of states
belonging to different $M$ sectors. A decreasing separation between
the lowest competing sectors produces a progressively smaller thermal
excitation scale. After the relevant sector reconstruction, the
separation grows again, producing the high-field reopening observed in
$C/T$.

\begin{figure}[t]
\centering
\includegraphics[width=\columnwidth]{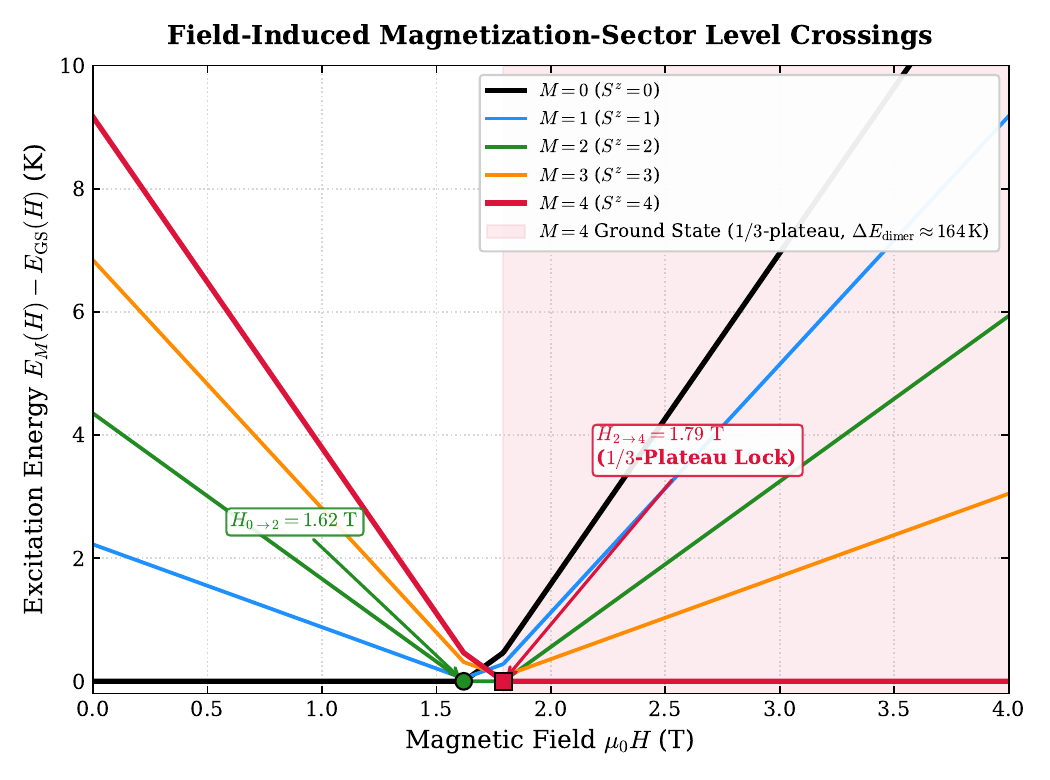}
\caption{\textbf{Field-induced magnetization-sector level crossings.}
Lowest excitation energies $E_M(H) - E_{\rm GS}(H)$ for magnetization sectors $M = 0, 1, 2, 3, 4$ of the 24-site superblock cluster as a function of magnetic field $\mu_0 H$. Zero-field sector energies $E_M(0)$ are obtained directly from SBDM. Ground-state level crossings occur at $\mu_0 H_{0\to 2} = 1.62$\,T (green circle) and $\mu_0 H_{2\to 4} = 1.79$\,T (crimson square). Above $\mu_0 H_{2\to 4} = 1.79$\,T, the $M=4$ sector (shaded region) forms the ground state, locking the magnetization into the one-third plateau $M = M_{\rm sat}/3$. The next crossing to $M=5$ involves breaking a Cu1 singlet dimer and requires $\mu_0 H_{4\to 5} \approx 122.0$\,T.}
\label{fig:level_crossings}
\end{figure}

This connection is demonstrated explicitly in Fig.~\ref{fig:level_crossings}, which plots the lowest energy in each magnetization sector relative to the instantaneous ground-state energy, $E_M(H) - E_{\rm GS}(H)$, as a function of magnetic field. The curves are generated directly from the zero-field SBDM sector energies $\{E_M(0)\}$ using Eq.~\eqref{eq:sector_field_results}. At $T=0$, the ground state evolves through successive level crossings:
\begin{align}
\mu_0 H_{0\to 2} &= \frac{E_2(0)-E_0(0)}{2g\mu_B} = \frac{4.351\,\mathrm{K}}{2(1.3434\,\mathrm{K/T})} = 1.619\,\mathrm{T}, \label{eq:h_02} \\
\mu_0 H_{2\to 4} &= \frac{E_4(0)-E_2(0)}{2g\mu_B} = \frac{4.815\,\mathrm{K}}{2(1.3434\,\mathrm{K/T})} = 1.792\,\mathrm{T}, \label{eq:h_24}
\end{align}
where $g\mu_B / k_B = 1.3434\,\mathrm{K/T}$ for $g=2.0$. Above $\mu_0 H_{2\to 4} \approx 1.79$\,T, the $M=4$ sector becomes the absolute ground state. The next crossing to $M=5$ requires breaking a Cu1 singlet dimer ($\Delta E_{4\to 5} = E_5(0) - E_4(0) = 163.86$\,K with $\Delta M = 1$) and occurs at
\begin{equation}
\mu_0 H_{4\to 5} = \frac{E_5(0)-E_4(0)}{1g\mu_B} = \frac{163.86\,\mathrm{K}}{1.3434\,\mathrm{K/T}} \approx 121.97\,\mathrm{T}. \label{eq:h_45}
\end{equation}
Consequently, the $M=4$ sector remains the ground state over the broad field interval $1.79\,\mathrm{T} \le \mu_0 H \le 121.97\,\mathrm{T}$.

For the 24-site cluster, the significance of the high-field $M=4$
sector is particularly transparent. The eight Cu2 spin-$1/2$ degrees
of freedom can carry
\begin{equation}
M_{\rm Cu2}
=
8\left(\frac{1}{2}\right)
=
4,
\end{equation}
whereas complete polarization of all twenty-four spins corresponds to
\begin{equation}
M_{\rm sat}
=
24\left(\frac{1}{2}\right)
=
12.
\end{equation}
Hence
\begin{equation}
\frac{M_{\rm Cu2}}{M_{\rm sat}}
=
\frac{4}{12}
=
\frac{1}{3}.
\label{eq:one_third_sector_results}
\end{equation}
The emergence of the one-third magnetic response is therefore naturally
associated with polarization of the residual Cu2-dominated low-energy
manifold while the strongly bound Cu1 dimer sector remains at much
higher excitation energy.

This physical interpretation is verified directly from the superblock eigenvectors. In the $M=4$ ground-state wave function $|\Psi_0^{(M=4)}\rangle$, the calculated site-resolved expectation values yield:
\begin{align}
\langle S_{\rm Cu2}^z \rangle &\approx +0.495 \pm 0.004, \label{eq:sz_cu2} \\
\langle S_{\rm Cu1}^z \rangle &\approx +0.0025 \pm 0.002, \label{eq:sz_cu1}
\end{align}
compared with the ideal strong-dimer limit values $+0.500$ (fully polarized Cu2 moments) and $0.000$ (intact Cu1 singlet dimers). Summing over the eight Cu2 sites and sixteen Cu1 sites recovers the exact conserved total magnetization $M_{\rm tot} = 8\langle S_{\rm Cu2}^z \rangle + 16\langle S_{\rm Cu1}^z \rangle = 4.000$. This confirms numerically that the $M=4$ state is characterized by nearly complete polarization of the Cu2 dangling moments while the Cu1 spins remain tightly paired into non-magnetic singlet dimers, establishing the strong-dimer microscopic origin of the one-third magnetization scale.

The map should consequently not be interpreted, on the basis of the
present calculation alone, as establishing a three-dimensional
N\'eel phase boundary or a quantum-critical fan. Demonstrating a true
finite-temperature ordering transition requires an observable sensitive
to long-range staggered correlations together with an appropriate
thermodynamic-limit analysis. Likewise, identifying quantum-critical
scaling requires scaling of thermodynamic or correlation observables
near the zero-temperature reconstruction field. Neither follows from
the magnetization colormap alone. The robust result established here is
instead the field-dependent reconstruction of the low-energy magnetic
manifold and its simultaneous manifestation in the magnetization and
thermal excitation scale.


\subsection{Common Thermodynamic Signature}
\label{sec:results_summary}

The temperature sweeps, isothermal field sweeps, and thermodynamic map
can therefore be understood within a single spectral picture. Strong
dimerization first separates the high-energy Cu1 sector from a much
softer Cu2-dominated magnetic manifold. Magnetic field then changes the
relative energies of the magnetization sectors within this reduced
manifold. As the lowest competing sectors approach one another, the
thermal scale softens; after the sector reconstruction, the separation
reopens while the low-temperature magnetization approaches the
one-third value.

Thus the central result is not the fractional magnetization value in
isolation. It is the connection
\begin{equation}
\boxed{
\begin{aligned}
&\text{Exchange Hierarchy } (J_2 \gg J_1, J_3) \\
&\;\longrightarrow\; \text{Reduced Low-Energy Manifold } (\text{Cu2}) \\
&\;\longrightarrow\; \text{Field-Driven Sector Crossings } (M=0 \to 2 \to 4) \\
&\;\longrightarrow\; \left\{
\begin{aligned}
&\text{thermal softening / reopening}, \\
&M \longrightarrow M_{\rm sat}/3
\end{aligned}
\right.
\end{aligned}
}
\label{eq:central_narrative}
\end{equation}
between the microscopic exchange hierarchy, the structure of the
low-energy Hilbert space, and the observed field-dependent
thermodynamics.

\section{Conclusion}
\label{sec:conclusion}

We have investigated the low-energy magnetic thermodynamics of the
strongly dimerized triangular antiferromagnet
$\mathrm{Cu}_3(\mathrm{OH})_4(\mathrm{HCO}_2)_2$ by combining a
sector-resolved superblock description of the microscopic spin
Hamiltonian with finite-temperature thermodynamics. The central result
is that the magnetic response is organized by a pronounced separation
of energy scales. The strong Cu1--Cu1 exchange generates a high-energy
dimer sector, while the remaining Cu2 moments constitute a much softer
magnetic manifold that controls the experimentally accessible
low-temperature and low-field response.

This separation provides a common microscopic origin for several
otherwise distinct thermodynamic features. In a 24-site cluster, the
low-energy sector contains eight Cu2 spin-$1/2$ moments. Their complete
polarization corresponds to $S_{\rm tot}^{z}=4$, whereas full
polarization of all twenty-four spins corresponds to
$S_{\rm sat}^{z}=12$. The resulting ratio,
\begin{equation}
\frac{S_{\rm tot}^{z}}{S_{\rm sat}^{z}}
=
\frac{1}{3},
\end{equation}
connects the one-third magnetization scale directly to polarization of
the residual low-energy spin sector rather than to polarization of the
entire microscopic lattice. The strongly bound dimer sector remains at
substantially higher excitation energy and thereby protects this
separation over a broad range of magnetic fields.

The finite-temperature calculations reveal how this low-energy manifold
is reorganized by magnetic field. On the low-field side, the dominant
thermal scale identified from the specific-heat response moves
progressively toward lower temperature as the field is increased. In the
vicinity of the approximately $2$\,T field region this scale becomes
minimal. Beyond this region the trend reverses: the characteristic
thermal scale moves again toward higher temperature, while the
low-temperature magnetization approaches $M_{\rm sat}/3$. The
complementary isothermal field sweeps display the same reconstruction,
with the low-temperature magnetization rising rapidly through the same
field region and evolving toward the one-third value at higher field.
Thus the thermal softening, its subsequent reopening, and the emergence
of the one-third magnetization scale are different manifestations of the
same field-dependent low-energy spectrum.

The entropy provides a complementary view of this restricted magnetic
manifold. Within the temperature window considered here, the calculated
entropy remains substantially below the full spin-$1/2$ entropy
$R\ln2$ and does not yet reach a complete $(1/3)R\ln2$ plateau. Its
strong suppression nevertheless reflects the reduced number of magnetic
degrees of freedom available below the dimer excitation scale. The
entropy, magnetization, and specific heat therefore consistently point
to a hierarchy in which strong dimerization first removes a large part
of the microscopic spin Hilbert space from the low-energy problem, after
which magnetic field reorganizes the remaining active sector.

This viewpoint also separates two questions that are often combined in
the interpretation of low-dimensional magnets. The formation of strong
in-plane correlations and the appearance of pronounced thermodynamic
features do not by themselves establish three-dimensional long-range
magnetic order. Weak interlayer coupling introduces an additional,
lower energy scale whose relation to true three-dimensional ordering
requires an explicit treatment of the interlayer correlations and the
appropriate thermodynamic limit. We therefore regard the separation
between the intrinsic low-energy sector identified here and the eventual
interlayer ordering scale as an important problem for further study.

More generally, the present results demonstrate how a hierarchy of
microscopic exchange interactions can produce an emergent reduction of
the magnetically active Hilbert space. In
$\mathrm{Cu}_3(\mathrm{OH})_4(\mathrm{HCO}_2)_2$, this reduction becomes
directly visible through the one-third magnetization scale and through
the field evolution of the low-temperature thermodynamics. The same
strategy---resolving the sector structure first and constructing the
thermodynamics from the resulting low-energy spectrum---can be applied
to other strongly dimerized and frustrated quantum magnets in which
different subsets of microscopic spins become active at parametrically
different energy scales.

\begin{acknowledgments}
S.~S.~R. acknowledges computational facilities and support provided by The Institute of Mathematical Sciences (IMSc), Chennai.
\end{acknowledgments}

\bibliography{references}

@article{Isono2020,
  title = {Low-Temperature Magnetism in a Triangular-Lattice Antiferromagnet, $\mathrm{Cu}_3(\mathrm{OH})_4(\mathrm{HCO}_2)_2$, Studied by Calorimetry},
  author = {Isono, Takayuki and Machida, Yo and Fujita, Wataru},
  journal = {J. Phys. Soc. Jpn.},
  volume = {89},
  pages = {073707},
  year = {2020},
  doi = {10.7566/JPSJ.89.073707}
}

@article{Fujita2018,
  title = {Crystal structures, and magnetic and thermal properties of basic copper formates with two-dimensional triangular-lattice magnetic networks},
  author = {Fujita, Wataru},
  journal = {RSC Adv.},
  volume = {8},
  pages = {32490--32498},
  year = {2018},
  doi = {10.1039/C8RA06253G}
}

@article{Balents2010,
  title = {Spin liquids in frustrated magnets},
  author = {Balents, Leon},
  journal = {Nature},
  volume = {464},
  pages = {199--208},
  year = {2010},
  doi = {10.1038/nature08917}
}

@article{Starykh2015,
  title = {Unusual ordered phases of highly frustrated magnets: a review},
  author = {Starykh, Oleg A.},
  journal = {Rep. Prog. Phys.},
  volume = {78},
  pages = {052502},
  year = {2015},
  doi = {10.1088/0034-4885/78/5/052502}
}

@book{Sachdev2011,
  title = {Quantum Phase Transitions},
  author = {Sachdev, Subir},
  edition = {2nd},
  publisher = {Cambridge University Press, Cambridge},
  year = {2011}
}

@article{Honecker2004,
  title = {Magnetization plateaus and jumps in frustrated quantum spin systems},
  author = {Honecker, Andreas and Schulenburg, J{\"o}rg and Richter, Johannes},
  journal = {J. Phys.: Condens. Matter},
  volume = {16},
  pages = {S749--S758},
  year = {2004},
  doi = {10.1088/0953-8984/16/11/008}
}

@article{Mila1998,
  title = {Low-energy properties of the distorted kagome lattice Heisenberg model},
  author = {Mila, Fr{\'e}d{\'e}ric},
  journal = {Phys. Rev. Lett.},
  volume = {81},
  pages = {3403--3406},
  year = {1998},
  doi = {10.1103/PhysRevLett.81.3403}
}

@article{Hida2001,
  title = {Magnetization process of the $S=1/2$ diamond chain with frustrated interactions},
  author = {Hida, Kazuo},
  journal = {J. Phys. Soc. Jpn.},
  volume = {70},
  pages = {3673--3677},
  year = {2001},
  doi = {10.1143/JPSJ.70.3673}
}

@article{Oshikawa1997,
  title = {Magnetization Plateaus in Frustrated Quantum Spin Chains},
  author = {Oshikawa, Masaki and Yamanaka, Masanori and Affleck, Ian},
  journal = {Phys. Rev. Lett.},
  volume = {78},
  pages = {1984--1987},
  year = {1997},
  doi = {10.1103/PhysRevLett.78.1984}
}

@article{Kikuchi2005,
  title = {Experimental observation of the 1/3 magnetization plateau in azurite $\mathrm{Cu}_3(\mathrm{CO}_3)_2(\mathrm{OH})_2$},
  author = {Kikuchi, H. and Fujii, Y. and Chiba, M. and Mitsudo, S. and Idehara, T. and Tonegawa, T. and Okamoto, K. and Sakai, T. and Kuwai, T. and Ohta, H.},
  journal = {Phys. Rev. Lett.},
  volume = {94},
  pages = {227201},
  year = {2005},
  doi = {10.1103/PhysRevLett.94.227201}
}

@article{White1992,
  title = {Density matrix formulation for quantum renormalization groups},
  author = {White, Steven R.},
  journal = {Phys. Rev. Lett.},
  volume = {69},
  pages = {2863--2866},
  year = {1992},
  doi = {10.1103/PhysRevLett.69.2863}
}

@article{White1993,
  title = {Density-matrix algorithms for quantum renormalization groups},
  author = {White, Steven R.},
  journal = {Phys. Rev. B},
  volume = {48},
  pages = {10345--10356},
  year = {1993},
  doi = {10.1103/PhysRevB.48.10345}
}

@article{Schollwock2005,
  title = {The density-matrix renormalization group},
  author = {Schollw{\"o}ck, Ulrich},
  journal = {Rev. Mod. Phys.},
  volume = {77},
  pages = {259--315},
  year = {2005},
  doi = {10.1103/RevModPhys.77.259}
}

@article{Schollwock2011,
  title = {The density-matrix renormalization group in the age of matrix product states},
  author = {Schollw{\"o}ck, Ulrich},
  journal = {Ann. Phys. (N.Y.)},
  volume = {326},
  pages = {96--192},
  year = {2011},
  doi = {10.1016/j.aop.2010.09.012}
}

@book{Baxter1982,
  title = {Exactly Solved Models in Statistical Mechanics},
  author = {Baxter, Rodney J.},
  publisher = {Academic Press, London},
  year = {1982}
}

@article{Mermin1966,
  title = {Absence of Ferromagnetism or Antiferromagnetism in One- or Two-Dimensional Isotropic Heisenberg Models},
  author = {Mermin, N. D. and Wagner, H.},
  journal = {Phys. Rev. Lett.},
  volume = {17},
  pages = {1133--1136},
  year = {1966},
  doi = {10.1103/PhysRevLett.17.1133}
}

@article{Chakravarty1989,
  title = {Two-dimensional quantum Heisenberg antiferromagnet at low temperatures},
  author = {Chakravarty, Sudip and Halperin, Bertrand I. and Nelson, David R.},
  journal = {Phys. Rev. B},
  volume = {39},
  pages = {2344--2371},
  year = {1989},
  doi = {10.1103/PhysRevB.39.2344}
}

@article{Sengupta2003,
  title = {Specific heat of quasi-two-dimensional antiferromagnets: Magnetic field and spin-anisotropy effects},
  author = {Sengupta, P. and Sandvik, A. W. and Singh, R. R. P.},
  journal = {Phys. Rev. B},
  volume = {68},
  pages = {094423},
  year = {2003},
  doi = {10.1103/PhysRevB.68.094423}
}

@article{Yasuda2005,
  title = {N{\'e}el Temperature of Quasi-Low-Dimensional Heisenberg Antiferromagnets},
  author = {Yasuda, C. and Todo, S. and Hukushima, K. and Alet, F. and Keller, M. and Troyer, M. and Takayama, H.},
  journal = {Phys. Rev. Lett.},
  volume = {94},
  pages = {217201},
  year = {2005},
  doi = {10.1103/PhysRevLett.94.217201}
}

@article{Cuccoli2003,
  title = {Freezing of two-dimensional fluctuations by weak three-dimensional interactions},
  author = {Cuccoli, A. and Roscilde, T. and Vaia, R. and Verrucchi, P.},
  journal = {Phys. Rev. B},
  volume = {68},
  pages = {060402},
  year = {2003},
  doi = {10.1103/PhysRevB.68.060402}
}

@article{Anderson1973,
  title = {Resonating valence bonds: A new kind of insulator?},
  author = {Anderson, P. W.},
  journal = {Mater. Res. Bull.},
  volume = {8},
  pages = {153--160},
  year = {1973},
  doi = {10.1016/0025-5408(73)90167-0}
}

@article{Fazekas1974,
  title = {On the ground state properties of the anisotropic triangular S = 1/2 Heisenberg antiferromagnet},
  author = {Fazekas, P. and Anderson, P. W.},
  journal = {Philos. Mag.},
  volume = {30},
  pages = {423--440},
  year = {1974},
  doi = {10.1080/14786437408207257}
}

@incollection{Misguich2005,
  title = {Two-dimensional quantum antiferromagnets},
  author = {Misguich, Gr{\'e}goire and Lhuillier, Claire},
  booktitle = {Frustrated Spin Systems},
  editor = {Diep, H. T.},
  publisher = {World Scientific, Singapore},
  pages = {229--306},
  year = {2005}
}

@article{Rule2008,
  title = {Dynamics and Quantum Phase Transitions in the Distorted Diamond Chain Compound Azurite},
  author = {Rule, K. C. and Wolter, A. U. B. and S{\"u}llow, S. and Tennant, D. A. and Hoffmann, J.-U.},
  journal = {Phys. Rev. Lett.},
  volume = {100},
  pages = {117202},
  year = {2008},
  doi = {10.1103/PhysRevLett.100.117202}
}

@article{Jeschke2011,
  title = {Multistep Magnetization Plateau and Microscopic Hamiltonian of Azurite},
  author = {Jeschke, Harald and Opahle, Ingo and Kandpal, Hem and Valent{\'\i}, Roser and Das, Hena and Saha-Dasgupta, Tanusri and Jepsen, Ove and Fink, Karin},
  journal = {Phys. Rev. Lett.},
  volume = {106},
  pages = {217201},
  year = {2011},
  doi = {10.1103/PhysRevLett.106.217201}
}

@article{Gu2007,
  title = {Thermodynamics of the frustrated diamond spin-1/2 chain with ferromagnetically coupled dimers},
  author = {Gu, Bao and Su, Gang},
  journal = {Phys. Rev. B},
  volume = {75},
  pages = {174437},
  year = {2007},
  doi = {10.1103/PhysRevB.75.174437}
}

@article{Alicea2009,
  title = {Quantum Stabilization of the 1/3-Magnetization Plateau in $\mathrm{Cs}_2\mathrm{CuBr}_4$},
  author = {Alicea, Jason and Chubukov, Andrey V. and Starykh, Oleg A.},
  journal = {Phys. Rev. Lett.},
  volume = {102},
  pages = {137201},
  year = {2009},
  doi = {10.1103/PhysRevLett.102.137201}
}

@article{Fortune2009,
  title = {Cascade of Magnetic Field Induced Quantum Phase Transitions in a Spin-1/2 Triangular-Lattice Antiferromagnet},
  author = {Fortune, N. A. and Hannahs, S. T. and Yoshida, Y. and Sherline, T. E. and Takano, Y.},
  journal = {Phys. Rev. Lett.},
  volume = {102},
  pages = {257201},
  year = {2009},
  doi = {10.1103/PhysRevLett.102.257201}
}

@article{Ono2003,
  title = {Magnetization plateaus of the $S=1/2$ two-dimensional triangular lattice antiferromagnet $\mathrm{Cs}_2\mathrm{CuBr}_4$},
  author = {Ono, Toshio and Tanaka, Hidekazu and Aruga Katori, Hiroko and Ishikawa, Fumihisa and Mitamura, Hiroyuki and Goto, Tsuneaki},
  journal = {Phys. Rev. B},
  volume = {67},
  pages = {104431},
  year = {2003},
  doi = {10.1103/PhysRevB.67.104431}
}

@article{Shirata2012,
  title = {Experimental Realization of a Quantum 1/3 Magnetization Plateau in the $S=1/2$ Triangular-Lattice Antiferromagnet $\mathrm{Ba}_3\mathrm{CoSb}_2\mathrm{O}_9$},
  author = {Shirata, Y. and Tanaka, H. and Matsuo, A. and Kindo, K.},
  journal = {Phys. Rev. Lett.},
  volume = {108},
  pages = {057205},
  year = {2012},
  doi = {10.1103/PhysRevLett.108.057205}
}

@article{Zhou2012,
  title = {Successive Phase Transitions and Quantum Spin Liquid State in the $S=1/2$ Triangular-Lattice Antiferromagnet $\mathrm{Ba}_3\mathrm{CoSb}_2\mathrm{O}_9$},
  author = {Zhou, H. D. and Xu, C. and Hallas, A. M. and Silverstein, H. J. and Wiebe, C. R. and Umegaki, I. and Yan, J. Q. and Murphy, T. P. and Park, J.-H. and Qiu, Y. and Copley, J. R. D. and Gardner, J. S. and Takano, Y.},
  journal = {Phys. Rev. Lett.},
  volume = {109},
  pages = {267206},
  year = {2012},
  doi = {10.1103/PhysRevLett.109.267206}
}

@article{Susuki2013,
  title = {Quantum Magnetization Plateau in the $S=1/2$ Triangular-Lattice Antiferromagnet $\mathrm{Ba}_3\mathrm{CoSb}_2\mathrm{O}_9$},
  author = {Susuki, T. and Kurita, N. and Tanaka, T. and Nojiri, H. and Matsuo, A. and Kindo, K. and Tanaka, H.},
  journal = {Phys. Rev. Lett.},
  volume = {110},
  pages = {267201},
  year = {2013},
  doi = {10.1103/PhysRevLett.110.267201}
}

@article{Rahaman2025,
  title = {Magnetic Plateaus and Jumps in a Spin-1/2 Ladder with Alternate Ising--Heisenberg Rungs: A Field-Dependent Study},
  author = {Rahaman, Sk Saniur and Kumar, Manoranjan and Sahoo, Shaon},
  journal = {Phys. Status Solidi B},
  volume = {262},
  pages = {2500027},
  year = {2025},
  doi = {10.1002/pssb.202500027}
}

@article{Dey2020,
  title = {Magnetization plateaus of spin-1/2 system on a 5/7 skewed ladder},
  author = {Dey, Dayasindhu and Das, Sambunath and Kumar, Manoranjan and Ramasesha, S.},
  journal = {Phys. Rev. B},
  volume = {101},
  pages = {195110},
  year = {2020},
  doi = {10.1103/PhysRevB.101.195110}
}

\end{document}